\documentclass{ws-procs975x65}

\begin{document}

\title{$T^{'}$ and the Cabibbo Angle}

\author{Paul H. Frampton}

\address{Department of Physics and Astronomy,
University of North Carolina, NC 27599\\
$^*$E-mail: frampton@physics.unc.edu}

\begin{abstract}
The use of the binary tetrahedral group ($T^{'}$) as
flavor symmetry is discussed. I emphasize the 
CKM quark and PMNS neutrino mixings.
I present a novel formula for the Cabibbo angle.
\end{abstract}


\section{Introduction on renormalizability}

In particle theory phenomenology, model building fashions
vary with time and because the present lack of data
(soon to be compensated by the Large Hadron Collider)
does not allow discrimination between models some fashions
develop a life of their own. In the present talk 
we take the apparently retrogressive step of imposing
the requirement of renormalizability, as holds
for quantum electrodynamics (QED), quantum chromodynamics (QCD)
and the standard electroweak model, to show that
non-abelian flavor symmetry becomes then
much more restrictive and predictive. In a specific model
we show that a normal neutrino mass hierarchy is
strongly favored over an inverted hierarchy.

For several years now there has been keen interest in the use
of $A_4$ as a finite flavor symmetry in the lepton
sector, especially neutrino mixing. In particular, the empirically approximate
tribimaximal mixing of the three neutrinos can
be predicted. It is usually stated that either normal or
inverted neutrino mass spectrum can be predicted.

We revisit these two questions in a
minimal $A_4$ framework with
only one $A_4$-{\bf 3} of Higgs doublets
coupling to neutrinos and permitting
only renormalizable couplings. For such a minimal model
there is more predicitivity regarding neutrino masses.

Although the standard model was originally discovered
using the criterion of renormalizability, it is sometimes
espoused that renormalizability
is not prerequisite in an effective lagrangian.
Nevertheless, imposing renormalizability in
the present case is more sensible because it
does render the model far more predictive
by avoiding the many additional parameters
associated with higher-order irrelevant operators.
Our choice of Higgs sector also
minimizes the number of free parameters.

It is sufficiently important to emphasize the concept
that every result mentioned in this talk would be
impossible without imposing renormalizability.

Although it has been fruitful
in low-energy QCD, heavy-quark effective theory
and technicolor this idea is inappropriate to
fundamental model building in particle phenomenology.

\section{\it $A_4$ symmetry}

The group $A_4$ is the order g=12 symmetry of a
regular tetrahedron $T$ and is a subgroup of the
rotation group $SO(3)$. $A_4$ has irreducible representations
which are three singlets $1_1, 1_2, 1_3$ and a triplet $3$.
In the embedding $A_4 \subset SO(3)$ the {\bf 3} of $A_4$
is identified with the adjoint {\bf 3} of $SO(3)$.

Since the only Higgs doublets coupling to neutrinos
in our model are in a {\bf 3} of $A_4$, it is very useful
to understand geometrically the three components
of a {\bf 3}.

A regular tetrahedron has four vertices, four faces and
six edges. Straight lines joining the midpoints of
opposite edges pass through the centroid and form
a set of three orthogonal axes. Regarding the regular
tetrahedron as the result of cutting off the four odd
corners from a cube, these axes are parallel to the
sides of the cube. With respect to the regular
tetrahedron, a vacuum expectation value (VEV)
of the {\bf 3} such as $<{\bf 3}> = v (1, 1, -2)$,
as will be used,clearly breaks SO(3) to U(1) and
correspondingly $A_4$ to $Z_2$, since it requires
a rotation by $\pi$ about the 3-axis to restore
the tetrahedron.

At the same time, we can understand the appearance
of tribimaximal mixing with matrix

\begin{equation}
U_{TBM} = \left(
\begin{array}{ccc}
- \sqrt{\frac{1}{6}} & -\sqrt{\frac{1}{6}} & \sqrt{\frac{2}{3}} \\
\sqrt{\frac{1}{3}} & \sqrt{\frac{1}{3}} & \sqrt{\frac{1}{3}} \\
\sqrt{\frac{1}{2}} & - \sqrt{\frac{1}{2}} & 0
\end{array}
\right),
\label{TBM}
\end{equation}

\noindent and our definitions are such that the ordering $\nu_{1, 2, 3}$
and $\nu_{\tau, \mu, e}$ satisfy

\begin{equation}
\left( \begin{array}{c} \nu_1 \\ \nu_2 \\ \nu_3 \end{array} \right)
= U_{TBM}
\left( \begin{array}{c} \nu_{\tau} \\ \nu_{\mu} \\ \nu_e \end{array} \right)
\end{equation}

Assuming no CP violation, the Majorana matrix $M_{\nu}$ is real and symmetric and
therefore of the form

\begin{equation}
M_{\nu} = \left(
\begin{array}{ccc}
A & B & C \\
B & D & F \\
C & F & E
\end{array}
\right)
\label{ABCDEF}
\end{equation}

and is related to the diagonalized
form by

\begin{equation}
M_{diag} = \left( \begin{array}{ccc}
m_1 & 0 & 0 \\
0 & m_2 & 0 \\
0 & 0 & m_3
\end{array}
\right) = U_{TBM} M_{\nu} U_{TBM}^{T}.
\label{diag}
\end{equation}

Substituting Eq.(\ref{TBM}) into Eq.(\ref{diag}) shows
that $M_{\nu}$ must be of the general form in terms
of real parameters $A, B, C$:
\begin{equation}
M_{\nu} = \left(
\begin{array}{ccc}
A & ~~~ B & C \\
B & ~~~ A & C \\
C & ~~~ C & ~~~ A + B -C
\end{array}
\right),
\label{ABC}
\end{equation}

which has eigenvalues

\begin{eqnarray}
m_1 &=& (A + B - 2C) \nonumber  \\
m_2 &=& (A + B + C) \nonumber \\
m_3 &=& (A - B).
\label{masses}
\end{eqnarray}

The observed mass spectrum corresponds approximately
to $|m_1| = |m_2|$ which requires either $C=0$ or $C=2(A+B)$.
For a normal hierarchy, $(A+B)=0$ and $C=0$. For an inverted hierarchy
$A=B$ and $C=0$ or $C=4A$.

Now we study our minimal $A_4$ model to examine the occurrence
of the Majorana matrix Eq.(\ref{ABC}) and the eigenvalues
Eq.(\ref{masses}).

\section{Minimal $A_4$ model}

We assign the leptons to $(A_4, Z_2)$ irreps as follows

\begin{equation}
\begin{array}{ccc}
\left. \begin{array}{c}
\left( \begin{array}{c} \nu_{\tau} \\ \tau^- \end{array} \right)_{L} \\
\left( \begin{array}{c} \nu_{\mu} \\ \mu^- \end{array} \right)_{L} \\
\left( \begin{array}{c} \nu_e \\ e^- \end{array} \right)_{L}
\end{array} \right\}
L_L  (3, +1)  &
\begin{array}{c}
~ \tau^-_{R}~ (1_1, -1)   \\
~ \mu^-_{R} ~ (1_2, -1) \\
~ e^-_{R} ~ (1_3, -1)  \end{array}
&
\begin{array}{c}
~ N^{(1)}_{R} ~ (1_1, +1) \\
~ N^{(2)}_R ~ (1_2, +1) \\
~ N^{(3)}_{R} ~ (1_3, +1).\\  \end{array}
\end{array}
\end{equation}

The lepton lagrangian is

\begin{eqnarray}
{\cal L}^{(leptons)}_Y
&=&
\frac{1}{2} M_1 N_R^{(1)} N_R^{(1)} + M_{23} N_R^{(2)} N_R^{(3)} \nonumber \\
& & + \left[
Y_{1} \left( L_L N_R^{(1)} H_3 \right) + Y_{2} \left(  L_L N_R^{(2)}  H_3 \right)
\right.  \nonumber \\
& & \left. + Y_{3}
\left( L_L N_R^{(3)} H_3 \right)  \right.  \nonumber \\
& & \left. +
Y_\tau \left( L_L \tau_R H'_3 \right)
+ Y_\mu  \left( L_L \mu_R  H'_3 \right) \right.  \nonumber \\
& & \left. +
Y_e \left( L_L e_R H'_3 \right)
\right]
+
{\rm h.c.}
\end{eqnarray}

\noindent
where $SU(2)$-doublet Higgs scalars are in $H_3(3, +1)$ and $H_3^{'}(3, -1)$.

\noindent
The charged lepton masses originate from
$<H_3^{'}> =
(\frac{m_{\tau}}{Y_{\tau}},\frac{m_{\mu}}{Y_{\mu}},\frac{m_{e}}{Y_{e}}) $
and are, to leading order, disconnected from the neutrino
masses if we choose a flavor basis where the charged leptons are mass eigenstates.
The $N_{R}^{i}$ masses break the $L_{\tau} \times L_{\mu} \times L_e$ symmetry but change the charged lepton masses
only by very small amounts $\propto  Y^2m_i/M_N$ at one-loop level.

\noindent
The right-handed neutrinos have mass matrix

\begin{equation}
M_N =
\left(
\begin{array}{ccc}
M_1 & 0 & 0 \\
0 & 0 & M_{23} \\
0 & M_{23} & 0
\end{array}
\right).
\label{MN}
\end{equation}

We take the VEV of the scalar $H_3$ to be

\bigskip

\begin{equation}
<H_3> = (V_1, V_2, V_3),
\label{vev}
\end{equation}

whereupon the Dirac matrix is

\begin{equation}
M_D =
\left(
\begin{array}{ccc}
Y_{1} V_1 & ~~~ Y_{2} V_3 & ~~~ Y_3 V_2 \\
Y_1 V_3 & ~~~ Y_2 V_2 & ~~~ Y_3 V_1 \\
Y_1 V_2 & ~~~ Y_2 V_1 & ~~~ Y_3 V_3
\end{array}
\right).
\label{MD}
\end{equation}

\noindent The Majorana mass matrix $M_{\nu}$ is given by

\begin{equation}
M_{\nu} = M_D M_N^{-1} M_D^{T}.
\end{equation}

Technical details are provided in arXiv:0806.1707.

Our conclusion is that 
the $A_4$ model in a minimal form does favor the normal hierarchy.
We have considered a more restrictive model based on $A_4$ than
previously considered. The theory has been required
to be renormalizable and the Higgs scalar
content is the minimum possible.
We have required that the neutrino mixing matrix be
of the tribimaximal form. We then find that the masses
for the neutrinos are highly constrained and
can be in a normal, not inverted hierarchy.

Most, if not all, previous $A_4$ models
in the literature
permit higher-order irrelevant non-renormalizable
operators and their concomitant proliferation
of parameters and hence allow a wide variety if
possiblities
for the neutrino masses.
We believe the renormalizabilty condition is sensible
for these flavor symmetries because of the higher predictivity.

The next step which is the subject of the rest
of this talk is whether
the present renornmalizable $A_4$ model can be extended
to a renormalizable $T^{'}$ model. It is necessary but
not
sufficent condition for this that a successful
renormalizable $A_4$ model, as presented here, exists.

\section{$T^{'}$ symmetry.}

The first use of the binary tetrahedral
group $T{'}$ in particle physics was \cite{Yang}
by
Case, Karplus and Yang in 1956
who were motivated to consider gauging a finite
$T^{'}$ subgroup of $SU(2)$
in Yang-Mills
theory.
This led Fairbairn, Fulton and Klink (FFK) in 1964
to make an analysis\cite{FFK}
of $T^{'}$ Clebsch-Gordan coefficients
As a flavor symmetry, $T^{'}$ first
appeared \cite{FK1}in 1994 motivated by
the idea of representing the three quark families
with the third treated differently from
the first two.
Since $T^{'}$ is the double cover
of $A_4$, it was natural to suggest that
$T^{'}$ be employed to accommodate quarks
and simultaneously
the established $A_4$ model building
for tribimaximal neutrino mixing.

\noindent
We shall discuss such a $T^{'}$ model
with simplifications to emphasize the largest quark mixing, the
Cabibbo angle,
for which we shall derive an entirely new formula \cite{FKM,EFM} as an exact angle.

\noindent
Recall that charged lepton masses arise from the vacuum expectation value

\begin{equation}
<H_3^{'}> =
\left(\frac{m_{\tau}}{Y_{\tau}},\frac{m_{\mu}}{Y_{\mu}},\frac{m_{e}}{Y_{e}}
\right) = ( M_{\tau}, M_{\mu}, M_e )
\label{Hprime}
\end{equation}

\noindent
where $M_i \equiv m_i/Y_i$ ($i = \tau, \mu, e$).
Neutrino masses and mixings
come from the see-saw mechanism
and the VEV

\begin{equation}
<H_3> = V( 1, -2, 1)
\label{VEV}
\end{equation}

\noindent
We shall now promote $A_4$ to $T^{'}$ keeping renormalizability and including quarks.

\section{ Minimal $T^{'}$ model}

\noindent
The left-handed quark doublets \noindent $(t, b)_L, (c, d)_L, (u, d)_L$
are assigned under $(T^{'} \times Z_2)$ to

\begin{equation}
\begin{array}{cc}
\left( \begin{array}{c} t \\ b \end{array} \right)_{L}
~ {\cal Q}_L ~~~~~~~~~~~ ({\bf 1_1}, +1)   \\
\left. \begin{array}{c} \left( \begin{array}{c} c \\ s \end{array} \right)_{L}
\\
\left( \begin{array}{c} u \\ d \end{array} \right)_{L}  \end{array} \right\}
Q_L ~~~~~~~~ ({\bf 2_1}, +1)
\end{array}
\label{qL}
\end{equation}

\noindent
and the six right-handed quarks as

\begin{equation}
\begin{array}{c}
t_{R} ~~~~~~~~~~~~~~ ({\bf 1_1}, +1)   \\
b_{R} ~~~~~~~~~~~~~~ ({\bf 1_2}, +1)  \\
\left. \begin{array}{c} c_{R} \\ u_{R} \end{array} \right\}
{\cal C}_R ~~~~~~~~ ({\bf 2_3}, -1)\\
\left. \begin{array}{c} s_{R} \\ d_{R} \end{array} \right\}
{\cal S}_R ~~~~~~~~ ({\bf 2_2}, +1)
\end{array}
\label{qR}
\end{equation}

\noindent
We add only two new scalars $H_{1_1} (1_1, +1)$ and
$H_{1_3} (1_3, +1)$ whose VEVs

\bigskip

\begin{equation}
<H_{1_1}> = m_t/Y_t ~~~~ <H_{1_3}> = m_b/Y_b
\label{H13VEV}
\end{equation}

\noindent
provide the $(t, b)$ masses.
In particular, no  $T^{'}$ doublet
($2_1, 2_2, 2_3$) scalars have been added. This
allows a non-zero value only for $\Theta_{12}$. The other angles
vanish making the third family stable
\footnote{As we shall discuss non-vanishing $\Theta_{23}$ and $\Theta_{13}$ are
related to $(d, s)$ masses.}.

\noindent
The allowed quark Yukawa and mass terms are

\begin{eqnarray}
{\cal L}_Y^{(quarks)}
&=& Y_t ( \{{\cal Q}_L\}_{\bf 1_1}  \{t_R\}_{\bf 1_1} H_{\bf 1_1}) \nonumber \\
&&
+ Y_b (\{{\cal Q}_L\}_{\bf 1_1} \{b_R\}_{\bf 1_2} H_{\bf 1_3} ) \nonumber \\
&&
+ Y_{{\cal C}} ( \{ Q_L \}_{\bf 2_1} \{ {\cal C}_R \}_{\bf 2_3} H^{'}_{\bf 3})
\nonumber \\
&&
+ Y_{{\cal S}} ( \{ Q_L \}_{\bf 2_1} \{ {\cal S}_R \}_{\bf 2_2} H_{\bf 3})
\nonumber \\
&&
+ {\rm h.c.}
\label{Yquark}
\end{eqnarray}

\noindent
The use of $T^{'}$ singlets and doublets
\footnote{It is discrete anomaly free.
We thank the UF-Gainesville group for discussions.}
for quark families
in Eqs.(\ref{qL},\ref{qR})
permits the third family to differ
from the first two and thus make plausible the
mass hierarchies $m_t \gg m_b$,  $m_b > m_{c,u}$
and $m_b > m_{s,d}$.

\section{The Cabibbo angle}

\noindent
The nontrivial ($2 \times 2$) quark mass matrices
$(c, u)$ and $(s, d)$ will be respectively denoted by $U^{'}$
and $D^{'}$ and calculated using the $T^{'}$ Clebsch-Gordan
coefficients of Fairbairn, Fulton and Klink.
 Dividing out $Y_{{\cal C}}$ and $Y_{{\cal S}}$
 in  Eq.(\ref{Yquark}) gives $U$ and $D$
 matrices ($\omega = e^{i \pi/3}$)

\begin{equation}
U \equiv \left( \frac{1}{Y_{{\cal C}}}\right) U^{'}  = \left(\begin{array}{cc}
\sqrt{\frac{2}{3}} \omega^2 M_{\tau} & \frac{1}{\sqrt{3}} M_e \\
- \frac{1}{\sqrt{3}} \omega^2 M_e & \sqrt{\frac{2}{3}} M_{\mu}
\end{array}
\right)
\label{Umatrix}
\end{equation}

\begin{equation}
D \equiv \left( \frac{1}{Y_{{\cal S}}} \right)
D^{'}  = \left( \begin{array}{cc}
\frac{1}{\sqrt{3}} & - 2 \sqrt{\frac{2}{3}} \omega \\
\sqrt{\frac{2}{3}} & \frac{1}{\sqrt{3}} \omega
\end{array}
\right)
\label{Dmatrix}
\end{equation}

\noindent
\noindent Let us first consider $U$ of Eq.(\ref{Umatrix}).
Noting that $m_{\tau} > m_{\mu} \gg m_e$
we may simplify $U$ by setting the electron mass
to zero, $M_e = 0$. This renders $U$ diagonal leaving
free the c, u, $\tau$ and $\mu$ masses.
This leaves only the matrix $D$ in Eq.(\ref{Dmatrix})
which predicts both $\Theta_{12}$ and $(m_d^2/m_s^2)$.
The hermitian square
${\bf {\cal D}} \equiv D D^{\dagger}$
is

\begin{equation}
{\bf {\cal D}} \equiv D D^{\dagger} = \left(
\frac{1}{3} \right) \left(
\begin{array}{cc}  9 & - \sqrt{2}  \\
- \sqrt{2} & 3
\end{array}
\right)
\label{DDdagger}
\end{equation}

\noindent
which leads by diagonalization to a formula for the Cabibbo angle

\begin{equation}
\tan 2\Theta_{12} = \left( \frac{\sqrt{2}}{3} \right)
\label{Cabibbo}
\end{equation}

\noindent
or equivalently $\sin \Theta_{12} = 0.218..$
close to the experimental value
footnote{Experimental results are from PDG2008;
 see references therein.}
 $\sin \Theta_{12} \simeq 0.227$.

\noindent
Our result of an exact angle for $\Theta_{12}$ can be regarded
as on a footing
with the tribimaximal values for neutrino angles $\theta_{ij}$.

Note that the tribimaximal $\theta_{12}$
presently agrees with experiment within one standard
deviation ($1 \sigma$).  On the other hand,
our analagous exact angle for $\Theta_{12}$
differs from experiment already by $9 \sigma$ which is
probably a reflection of the fact
that the experimental accuracy for $\Theta_{12}$ is $\sim 0.5\%$
while that for $\theta_{12}$ is $\sim 6\%$.

\noindent 
It is thus very important
to acquire better experimental data on $\theta_{12}$,
$\theta_{23}$ and $\theta_{13}$
to detect their similar deviation from the exact angles
predicted by TBM.
Our result for $(m_d^2/m_s^2)$ from Eq.(\ref{DDdagger}) is exactly
$0.288..$ compared to the central experimental value
$\simeq 0.003$ in a simplified
model whose generalization to an extended
scalar sector including $T^{'}$ doublets can
avoid $\Theta_{23} = \Theta_{13} = 0$
and thereby change $(m_d^2/m_s^2)$
due to mixing of $(d, s)$
with the $b$ quark.

\noindent
This $T'\times Z_2$ extension of the standard model is an
first step to tying the quark and lepton sectors together,
providing calculability, and at the same time reducing the number
of standard model parameters. The ultimate goal would be to
understand the origin of this discrete symmetry. Since gauge
symmetries can break to discrete symmetries, and gauge symmetries
arise naturally from strings, perhaps there is a clever
construction of our model with its fundamental origin in string theory.

\section{Summary}

\noindent
Renormalizability and simplification of $(A_4 \times Z_2)$
then $(T^{'} \times Z_2)$ models lead to:

\noindent Cabibbo angle formula

\[
\tan 2\Theta_{12} = \left( \frac{\sqrt{2}}{3} \right)
\]

\end{document}